\documentstyle[aps,prl,multicol]{revtex}
\input {epsf.sty}
\begin{document}
\draft
\title{Exclusion Statistics in Classical Mechanics}
\author{T.H. Hansson$^{1}$, S.B. Isakov$^{2 \; \star}$, J.M.
Leinaas$^{2}$,  and  U. Lindstr{\"o}m$^{1}$}
\address{$^{1}$ Department of Physics, University of
Stockholm, Box 6730, S-11385 Stockholm, Sweden }
\address{ $^{2}$ Department of Physics,  University of Oslo,
P.O. Box 1048 Blindern, N-0316 Oslo, Norway }
\date{April 15, 2000}
\maketitle
\begin{abstract}
We present a general method to derive the classical mechanics of a system
of identical particles in a way that retains information about quantum
statistics. The resulting statistical mechanics can be interpreted as a
classical version of Haldane's exclusion statistics.
\end{abstract}
\vspace{-11pt}
\begin{multicols}{2}

\newcommand{\half}{\frac 1 2 }
\newcommand{\eg}{{\em e.g.} }
\newcommand{\ie}{{\em i.e.} }
\newcommand{\etc} {{\em etc.}}

\newcommand{\noi}{\noindent}
\newcommand{\etal}{{\em et al.\ }}
\newcommand{\cf}{{\em cf. }}

\newcommand{\dd}[2]{{\rmd{#1}\over\rmd{#2}}}
\newcommand{\pdd}[2]{{\partial{#1}\over\partial{#2}}}
\newcommand{\pa}[1]{\partial_{#1}}
\newcommand{\pref}[1]{(\ref{#1})}
\newcommand{\kah}{K\"ahler } 

\def\zb{\overline z}
\def\zpb{\overline z'}
\def\ad{a^{\dagger}}
\def\bz{{\bf z}}
\def\bzb{{\bf \zb}}
\def\bm{{\bf m}}

\newcommand{\bra}[1]{\langle #1 |}
\newcommand{\ket}[1]{|#1\rangle}
\newcommand{\bracket}[2]{\langle #1|#2 \rangle}
\newcommand{\dop}[1]{D(#1)}

\newcommand{\za}[1]{z_{#1}}
\newcommand{\zab}[1]{\overline z_{#1}}

\newcommand{\pza}[1]{\partial_{z_{#1}} }
\newcommand{\pzab}[1]{\partial_{\overline z_{#1}} }

\newcommand{\zz}[2]{\overline z_{#1}z_{#2}  }
\newcommand{\braket}[2]{\langle #1 | #2 \rangle}
\newcommand{\nN}{{\cal N}}

\newcommand{\jpa}[1]{J_{+}^{#1} }
\newcommand{\jma}[1]{J_{-}^{#1} }
\newcommand{\epp}{(1+\zb z)}
\newcommand{\eppa}[2]{(1+\zb_{i} z_{#2})}
\newcommand{\eppj}{(1+\frac {\zb z} {2j})}
\newcommand{\eppja}[2]{(1+\frac {2\zb_{i} z_{#2}} {j})}
\newcommand{\aab}[1]{(1+\overline #1 #1)}

\newcommand{\bea}{\begin{eqnarray}} 
\newcommand{\eea}{\end{eqnarray}}
\newcommand{\e}{\varepsilon} 
\newcommand{\D}{\partial}

\newcommand{\ee}{\end{eqnarray}}


\newcommand {\be}[1]{
      \begin{eqnarray} \mbox{$\label{#1}$}  }

Particle statistics enters quantum physics in two related but logically
distinct way. The first one is related to the symmetry of the wave
function, or more generally to phase factors associated with the exchange
of identical particles. The other is related to entropy, \ie to the
counting of quantum states, and is expressed through the Pauli exclusion
principle and the phenomenon of Bose condensation. Some years ago,
Haldane pointed out that it is possible to have a certain kind of quantum
statistics of the second type, so called exclusion statistics, without
any reference to wave functions or exchange factors
\cite{Haldane91}.  While the exchange phase factors characterizing
fermions, bosons or  (in 2 space dimensions) 
anyons\cite{Laidlaw71,Leinaas77} are intimately connected to quantum
mechanics, there is no logical reason for not having  effects of the
second type of statistics even in classical systems. In fact, as is well
known such effects must be put in by hand in order to avoid the Gibbs
paradox in classical statistical mechanics. 

In this letter we show that it is possible to formulate a classical
mechanics  that builds in the effects of quantum statistics at the 
Lagrangian level. The dynamics, \ie the equations of motion, will not
depend on the statistics parameter, but the counting of states, and thus
the statistical mechanics, will. In the classical description the
statistics is expressed through the occupation of phase space volume.
Thus, each new particle introduced in the system will reduce the
available phase space volume for the other particles, and the degree of
reduction defines the classical statistics parameter.

We first sketch a general formulation of the problem, and then present
some specific results for two cases, non-interacting charged particles in
a strong magnetic field and vortices in the Landau-Ginzburg-Chern-Simons
theory. Both these systems are of  interest for the quantum Hall effect.
In both  examples, we show that the resulting classical statistical
mechanics is a classical version of  Haldane's exclusion statistics. Our
examples are two-dimensional, but just as in the case of exclusion
statistics,  there is  no reason in principle for our construction not to
work in an arbitrary  dimension. Below we will only present the general
ideas and some of the main results.  A fuller account including
calculational details can be found in ref. \cite{hansson2000}.

Consider a general quantum system and a subset of states $\ket
{\psi_\bz}$, which is labelled by a set of complex coordinates
$\bz=\{z_1,z_2,...,z_N\}$. These may be the coordinates of a system of
(identical) particles or the coordinates of an $N$ soliton
configuration. We only assume that the wave function evolves smoothly
with a change of these coordinates, and that it is symmetric under an
interchange of any pair of the $N$ coordinates.  To define the
corresponding classical mechanics, consider the constrained system where
the evolution of the full quantum system is projected to the manifold,
{$\cal M$}, defined by  the (normalized) states
$\ket {\psi_\bz}$.    The Schr{\"o}dinger equation of the full system
can be derived from the Lagrangian, 
\be{glag1} 
L=i\hbar \bracket \psi {\dot \psi} - \bra\psi  H \ket\psi \; ,
\ee 
and the Lagrangian of the constrained system is obtained from this 
by restricting $\ket\psi$ to the subset of states $\ket{\psi_\bz}$.
Expressed in terms of the coordinates
$\bz$, it is of  the generic form,
\be{nlag3} L(\bz , \bzb ) =  A_{\zab i}
\dot{\zab i}  + A_{\za i}  \dot{\za i}  - V(\za i, \zab i) \; , 
\ee where $A_z$ is the Berry connection,
\be{gpot} A_{z_i}=i\hbar \braket {\psi_z} {\partial_{z_i}{\psi_x}}\; ,
\ee
$V$ is  the expectation value of the Hamiltonian in the state
$\ket {\psi_\bz}$. 

An important special case is when the
state vectors which define ${\cal M}$ are, up to normalization, analytic
functions of
$z_i$,
\be{gnorm}
\ket{\psi_\bz}=\nN(\bzb,\bz) \ket{\phi_\bz} \; ,
\ee 
where $\ket{\phi_\bz}$ denotes the analytic part of the state
vector, and
$\nN(\zb,z)$ is the normalization factor. The vector potentials are then
given by
\be{gpot2} A_{z_i}&=&-i\hbar\partial_{z_i}\ln{\overline\nN(\bzb,\bz)}\;
, \nonumber \\
\ee 
and 
 $\cal M$ is a K{\"a}hler manifold where the K{\"a}hler potential is
related in a simple way to the normalization factor,
\be{gkh2}
K(\bzb,\bz)=\hbar\ln |\nN (\bzb,\bz)|^{-2} \; .
\ee
The geometry of ${\cal M}$ is described by the field strength
\be{gfs3}
f_{\zb_iz_j}&=&\partial_{\zb_i}A_j-\partial_{z_j}A_{\bar i} 
= i\partial_{\zb_i}\partial_{z_j}K(z,\zb)\; ,
\ee
in terms of which the symplectic form, $\omega$,  and the metric, $ds^2$, takes
the forms  $\omega=-f_{\zb_iz_j}d\zb_i\wedge dz_j$ and 
$ds^2=-2i f_{\zb_iz_j}d\zb_idz_j$.
The importance of the symplectic form $\omega$
is that it determines the Poisson brackets, and thus,
together with the Hamiltonian, defines the classical mechanics 
 \cite{Jackiw1}. The Poisson bracket is
\be{brack}
\{A,B\}=f_{z_i\zb_j}^{-1} \,\partial_{z_i}A\,\partial_{\zb_j} B \; ,
\ee
and the equation of motion  can then be written as
\be{geom2}
\dot z_i=\{z_i,V\} \; .
\ee

To be more specific we now consider a system of charged 
particles moving in two dimensions
 in the presence of a strong magnetic field that restricts 
the available states to the lowest Landau level.
In this example we can explicitly derive the metric and 
symplectic form, and show that they can be obtained 
from a K{\"a}hler potential. For  calculations, it is
convenient to consider  particles moving on a sphere. 
On a unit sphere penetrated by $2j$ units of magnetic flux 
a particle with unit charge has a total angular momentum 
$J=j+L$, where $L$ is the orbital angular momentum, and the lowest Landau 
level corresponds to $L=0$ with a $2j+1$ degeneracy \cite{Haldane83}.
We shall use the notation of ref.~\cite{pere1} and define a coherent 
state by rotations of a minimum uncertainty reference state $\ket 
{ 0}$, 
\be{csdef}
\ket z = D(z)\ket {0} 
 =   e^{ z J_{+} } 
 e^{\eta J_{0}}  e^{-\zb  J_{-}}  \ket{0} \; .
\ee
Here the complex coordinate $z$ is defined via a stereographic 
projection, and the  rotation operators, $D(z)$, form a unitary
and irreducible representation of the rotation group, generated by
 $J_{m}$, $\eta = \ln(1+ \zb z )$, and 
 $\ket{0}$  is annihilated by $J_{+}$. 

Fully symmetrized and antisymmetrized states corresponding to fermions
and bosons  are given by 
\be{spsymcs}
\ket{\bz , \pm} = 
\nN (\bz , \bzb ) \frac 1 
 {\sqrt N!} \sum_{P}\eta_{P}^{\pm}  e^{ \za {i_{P}} \jpa i } 
 \ket{\bf{0}} \; ,
 \ee
with $\eta_{P}^{\pm}$ the appropriate sign for the permutation $P$.  The
normalizations of these states are readily  obtained from the properties 
of the $D(z)$:s, 
\be{spnorm2}
|\nN(\bz,\bzb)|^{-2} = \sum_{P}\eta_{P} \prod\limits_i\left(1 + {\zab {i_{P}} 
\za i}  \right)^{2j}
\; .
\ee
For the case of N coinciding bosons, $\za i = z$,   we immediately get
the following \kah potential
\be{coinbos}
K(z, \zb)  =N\hbar 2j\ln(1 + {\zb z}) \; ,
\ee
and the corresponding metric,  
$ds^2 =  {2N\hbar}/ {\left(1 + {\zb z}\right)^{2}} dz d\zb$, 
is just $N$ times that of a sphere. 

To assess the effect of statistics in the classical description we 
calculate the $N$-particle phase space volume.
Following Manton \cite{Manton93}, and Samols
\cite{Samols92} we can use  \pref{coinbos} to obtain 
the $N$-particle volume from the volume of
$N$ coinciding bosons, and the result is,
\be{bphs}
V_{B} = \frac 1 {N!} (A)^{N} \; ,
\ee
with $A$ as the volume of the single-particle space, 
\be{fundform}
A = h\int_{sph}\, \omega = h 2  j = \frac {\hbar4\pi R^{2}}
{l^{2}} = e\Phi\; ,
\ee
which is $h$ times the number of flux quanta  $\phi_{0}=h/e$ that
penetrate the sphere. Thus, for bosons the only effect of the
indistinguishability of the particles is the factor $1/N!$, and there is
no further reduction in phase space volume. 

One should note that the classical phase space defined as above is
everywhere a smooth manifold. This is different from the case when the
N-particle space is defined as a product of single particle spaces with
identification of equivalent configurations. In the latter case the
points corresponding to a coincidence of two particle positions are
singular. Also note that the factor $1/N!$ here appears naturally from
the geometry, not through any additional assumption about identification
of points.

For fermions a similar but technically more involved calculation can be
done. The resulting phase space volume is,
\be{fphs}
V_{F} = \frac 1 {N!} \left(A - (N-1)h  \right)^{N} \; .
\ee
Compared with the Bose case there is an additional reduction of phase
space. The available phase space for any  particular fermion is reduced
with an amount
$h$ by each of the other particles present in the system.  This can be
understood  as a {\em classical version of the Pauli exclusion
principle,} and is consistent with the usual semi-classical
interpretation of quantum mechanics, where each quantum state is
associated with a phase space volume
$h$. Note that there is a maximum number of particles allowed, $N=2j+1$,
in which case the phase space volume
\pref{fphs} vanishes. This corresponds to  all  states with zero 
orbital angular momentum  being filled, \ie to a filled lowest Landau
level, which is  an incompressible state.

The calculation can be done also in the case of anyons although there is
no explicit expression for the normalization constant  corresponding  to
\pref{spnorm2}. For details we again refer to \cite{hansson2000} and
only quote the  result,
\be{afphs}
V_{\nu} = \frac 1 {N!} \left(A - \nu (N-1)h \right)^{N} \; ,
\ee
where the exchange phase of the anyons is $\nu\pi$.

The expressions we have found above for the $N$-particle phase space
volume demonstrates a {\em classical fractional exclusion principle}. 
Thus, each new particle introduced in the system will find the available
volume reduced by $\alpha =\nu h$ relative to the previous one. The
quantity
$\alpha$, \ie the reduction in phase space volume, can be taken as
defining the classical statistics parameter of the particles. In the
present case it is simply the (dimensionless) quantum statistics
parameter
$\nu$ multiplied with Planck's constant $h$. 

We next consider a   classical field theory with soliton
solutions, namely  the Chern-Simons Ginzburg-Landau (CSGL) 
theory, originally introduced as a field theory for the quantum Hall
effect \cite{qhalleffth},
\be{oglcs1}
L=\int d^2x[i\hbar\phi^*D_0\phi &-&\frac{\hbar^2}{2m}|\vec
D\phi|^2-\frac{\lambda}{4}
(|\phi|^2-\rho_0)^2 \\
&+&\mu\hbar\epsilon^{\mu\nu\rho}a_\mu\partial_\nu a_\rho] \nonumber \; ,
\ee
where $\phi$ is a  complex matter field, $a_\mu$ a Chern-Simons
field, $m$  a mass parameter,
$\lambda$  the interaction strength,  $\rho_0$  the preferred
density of the system and $\mu$ a statistics parameter. 
For the original Laughlin quantum Hall states
described by the model the statistics parameter takes the values
$\mu=1/[4\pi(2k+1)]$.
This theory has  vortices (quasi-particles) as soliton solutions. In
a certain approximation the dynamics can be described in terms of vortex
coordinates alone, and a phase space description can be derived from the
full theory. Again it is possible to calculate the phase space volume 
corresponding to a $N$-vortex solution.

The vortex configurations can be  parameterized by a set of vortex
coordinates
$\bz=\{z_1,z_2,...,z_n\}$, just as the charged particles in a magnetic 
field discussed above.  The precise form of the multi-vortex
configurations for given coordinates is not known, but for critical
coupling ($\lambda=1$) the existence of $N$-vortex configurations with
arbitrary positions can be deduced \cite{Taubes80}. With the system
constrained to the manifold of $N$-vortex configurations, a classical
mechanics follows with a kinetic term for the $N$-vortex system
corresponding to a phase space with K{\"a}hler metric
\cite{hansson2000}.  By use of earlier results obtained by Manton for
the related relativistic abelian Higgs  model \cite{Manton93} we find
for the phase space volume of $N$ vortices,
\be{volume2} 
V_N={1\over{N!}}(A-4\pi\mu h (N-1))^N \; ,
\ee
where the classical statistics parameter, as determined by the reduction
in available phase space due to the presence of other vortices,  is
$\alpha=4\pi\mu h\equiv g h$. We can interpret $g$, the classical
parameter divided by $h$, as the dimensionless quantum statistics
parameter. The value
$g= 4\pi\mu$ agrees with the value of the statistics parameter as
determined from  Berry phase calculations with Laughlin wave functions
\cite{Arovas84},  or from the properties of vortices in the CSGL model
\cite{qhalleffth}.  

We now discuss the statistical mechanics of the classical systems  just 
derived. Both these systems have the special property that the energy
does not depend on the state, but only on the number of particles. This
means that the statistical mechanics is determined by the phase space
volume
$V_N$, which has been determined in the previous sections, and by the
energy
$E_N$. The classical partition function is simply the total number  of
states,
$ { V_{N}}/ h^{N}$ multiplied  with the Boltzmann factor, \ie 
\be{part}
Z_{N} =\frac { V_{N}} {h^{N}  } e^{-\beta E_{N}} \; ,
\ee
Using standard thermodynamics we get for the entropy
\be{ent}
S &=& N\ln(1 - \alpha\rho) + N\ln \frac A h - N\ln N + N \; ,
\label{pres2}
\ee
where  $\alpha=\nu h$ or $gh$, and where we have introduced the
classical phase space density $\rho =N/A$  and neglected the difference
between
$N$ and
$N-1$, which is  irrelevant in the thermodynamic limit. 

In  the systems we have considered the real two-dimensional space where 
the particles or vortices move is proportional to the phase space.
Defining the pressure as,
$P = -\left({\partial F} /{\partial A} \right)_T$,  where $A=V_{1}$ is
the phase space volume for a single particle,  we then get, by use of
standard thermodynamic relations, the equation of state 
\be{eq.state}
\beta P &=&  \rho /{(1 - \alpha \rho)}.
\ee
This expression  shows that there is a maximum density
$\rho =1/\alpha$ allowed by the system, which corresponds to an infinite
pressure and therefore to an incompressible state. For the phase space
volume this means $V_{N} = 0$, \ie there is no available phase space
volume for any new particle added to the system. For the anyon system
this situation corresponds to a completely filled Landau level. What is
unusual about this is that the blocking, which can be interpreted as
representing a generalized Pauli principle, shows up not only in the
quantum,  but also in the classical description of the system. 

Finally we show that the thermodynamics just derived can be viewed as a 
classical limit of Haldane exclusion statistics \cite{Haldane91}.

The statistical mechanics of particles with exclusion statistics can be
derived by assuming that the total energy can be written as a sum of
single-particle energies and using  the prescriptions for statistical
weight given by Haldane separately for each single-particle energy level
\cite{Isakov94,dVO94,Wu94}. The result for the entropy is
\be{exent}
S &=& \sum_{k} D_{k}  \{ [1+(1-g)n_{k}] \ln [1+(1-g)n_{k}] \\ &+& 
(1-gn_k) \ln (1-gn_{k}) - n_{k} \ln n_{k} \} \; , \nonumber
\ee 
where the sum runs over single-particle energy states, and $g$ is the 
exclusion statistics parameter. $D_{k}$ is the
degeneracy of  the $k$-th level and  the quantum distribution
function $n_{k}$ is the average occupation number of the state $k$. 

Since each quantum state occupies the phase space volume $h^{\cal D}$,
with $2{\cal D}$  the dimension of the single-particle phase space,  we
can relate $n$ and
$\rho$ in  the semiclassical limit by $n=\rho h^{\cal D}$. In the
Boltzmann limit,  $h \rightarrow 0$ and $n\rightarrow 0$, all 
dependence on $g$ in \pref{exent} goes away. If we, however,  define the
classical physics by the limit  $h \rightarrow  0$, $g \rightarrow
\infty$ and $gh^{\cal D} \rightarrow \alpha$, where 
$\alpha$ is interpreted as a classical statistics parameter
\pref{exent} gets the nontrivial  limit of
\be{cexent}
S = \sum_{k} D_{k}h^{\cal D}\left[ \rho_k\ln(1-\alpha\rho_k) -
\rho_k\ln(\rho_k h) +\rho_k 
\right]
\; .
\ee
If we further specialize to the case of  fully degenerate states in a
two-dimensional phase space,   where the sum is simply replaced by the
total number of  available single-particle states, $G = A/h$, and where
$\rho_k$ is replaced by $N/A$, we exactly regain \pref{ent}. This
demonstrates that the classical statistical mechanics discussed in the
previous section can be regarded as a special limit of exclusion
statistics, different from the Boltzmann limit.  Starting from the the
equation of state for exclusion statistics particles with  the same
energy, we can also derive \pref{eq.state} if we identify the the
physical volume ${\cal V}$ with the one  particle phase space volume
$A$. 

One can also obtain exact results when the particles move in an external
harmonic potential. Again one finds equivalence between the statistical
mechanics derived from the classical mechanics of identical particles
with a statistics parameter and  the statistical mechanics derived from
exclusion statistics in the classical limit discussed above
\cite{hansson2000}. 

Although the phase space in the examples discussed in this paper are
two-dimensional, there is no obvious reason for the construction not to
work in higher dimensions. One point of particular interest to study 
further is how the ``classical fermion'' theory discussed here can be
applied to systems of interacting fermions
moving in two or three dimensions.

We thank A. Karlhede for discussions and comments on the manuscript.
T. H. Hansson and U. Lindstr{\"o}m were supported by the Swedish Natural 
Science Research Council. S. Isakov acknowledges the support 
received through a NATO Science
Fellowship granted by the Norwegian Research Council, and also
appreciates the warm hospitality of NORDITA during his stay there in the
summer of 1999, where part of this work was done.

\noindent $^{\star}$ Present address: NumeriX Corporation,
546 Fifth Avenue, New York, NY 10036, USA

\vspace{-0.5cm}
\bibliographystyle{unsrt}

\vspace{0.5cm}
\end{multicols}
\end{document}